\documentclass[prd,aps,floats,floatfix,nofootinbib]{revtex4}
\usepackage{amsmath}
\usepackage{amsfonts}
\usepackage{graphicx}
\usepackage{slashed}
\usepackage{dcolumn}
\usepackage{bm}
\usepackage[dvips]{color}
\newcommand{\beq}{\begin{equation}}
\newcommand{\eeq}{\end{equation}}
\begin{document}
\newcommand{\vect}[1]{\overrightarrow{#1}}
\newcommand{\smbox}[1]{\mbox{\scriptsize #1}}
\newcommand{\tanbox}[1]{\mbox{\tiny #1}}
\newcommand{\vev}[1]{\langle #1 \rangle}
\newcommand{\Tr}[1]{\mbox{Tr}\left[#1\right]}
\newcommand{\cosb}{c_{\beta}}
\newcommand{\sinb}{s_{\beta}}
\newcommand{\tanb}{t_{\beta}}
\newcommand{\picwidth}{3.4in}

\title{Condensate Enhancement and $D$-Meson Mixing in Technicolor Theories}

\author{R. Sekhar Chivukula}
\email[]{sekhar@msu.edu}
\author{Elizabeth H. Simmons}
\email[]{esimmons@msu.edu}
\affiliation{Department of Physics,
Michigan State University, East Lansing, MI 48824, USA}
\date{\today}

\begin{abstract}
Since the pioneering work of Eichten and Lane it has been known that the scale 
of the interactions responsible for the generation of the strange-quark mass in 
extended technicolor theories  must, absent any ``GIM-like" mechanism for suppressing 
flavor-changing neutral currents, be greater than of order 1000 TeV. In this note 
we point out that the constraint from the neutral $D$-meson system is now equally strong, 
implying that the charm quark mass must also arise from flavor dynamics at a scale 
this high. We then quantify the degree to which the technicolor condensate must be 
enhanced in order to yield the observed quark masses, if the extended technicolor scale 
is of order 1000 TeV. Our results are intended to provide a framework in which to interpret
and apply the results of lattice studies of conformal strongly interacting gauge theories, 
and the corresponding numerical 
measurements of  the anomalous dimension of the mass operator  in candidate 
theories of ``walking" technicolor.
\end{abstract}

\maketitle

\section{Introduction}

Technicolor \cite{Weinberg:1979bn,Susskind:1978ms,Hill:2002ap} 
provides a dynamical mechanism for electroweak symmetry breaking in which
the weak interactions are spontaneously broken to electromagnetism via technifermion chiral symmetry
breaking (which is analogous to quark chiral symmetry breaking in QCD). While technicolor chiral
symmetry breaking alone is sufficient to generate the masses of the weak gauge bosons,
additional ``extended technicolor" (ETC) interactions \cite{Dimopoulos:1979es,Eichten:1979ah}
are required to couple the symmetry breaking sector to the quarks and leptons and thereby 
generate ordinary fermion masses. As noted by Eichten and Lane \cite{Eichten:1979ah},
however, the additional interactions introduced to generate ordinary fermion masses cannot
be flavor-universal, and would therefore also give rise to flavor-changing
neutral-current (FCNC) processes. In particular they showed that, absent any ``GIM-like" mechanism 
\cite{Chivukula:1987py,D'Ambrosio:2002ex,Appelquist:2004ai,Martin:2004ec} for suppressing 
flavor-changing neutral currents, the ETC scale associated with strange-quark mass generation
must be larger than of order $10^3$ TeV in order to avoid unacceptably large ($CP$-conserving) contributions to neutral
$K$-meson mixing. To obtain quark masses that are large enough therefore requires an
enhancement of the technifermion condensate over that expected naively
by scaling from QCD. Such an enhancement can occur in ``walking" technicolor theories
in which the gauge coupling runs very slowly \cite{Holdom:1981rm,Holdom:1984sk,Yamawaki:1985zg,Appelquist:1986an,Appelquist:1986tr,Appelquist:1987fc}\footnote{For some examples of proposed models of walking technicolor, see \protect\cite{Appelquist:1997fp} and \protect\cite{Sannino:2009za} and references therein.}, or in ``strong-ETC" theories in which
the ETC interactions themselves are strong enough to help drive technifermion chiral symmetry breaking
\cite{Appelquist:1988as,Takeuchi:1989qa,Miransky:1988gk,Fukano:2010yv,Matumoto:1989hf}.\footnote{It is also notable that walking technicolor
and strong-ETC theories are quite different from QCD, and may be far less constrained by precision
electroweak measurements \protect\cite{Peskin:1990zt,Peskin:1991sw,Hill:2002ap}.}

In this paper we update the bounds on ETC interactions derived from limits on flavor-changing neutral-currents. In particular
we show that the bound on the scale of ETC interactions 
arising from $D$-meson mixing is now as constraining as that arising
from $CP$-conserving contributions to $K$-meson mixing and, therefore, absent any mechanism for the suppression of
flavor-changing neutral-currents \cite{Chivukula:1987py,D'Ambrosio:2002ex,Appelquist:2004ai,Martin:2004ec}, the 
ETC scale associated with {\it charm}-quark mass generation
must also be larger than of order $10^3$ TeV. Since the charm quark is so much heavier than the strange quark, requiring an ETC model to produce $m_c$ from interactions at a scale of over $1000$ TeV is a significantly stronger constraint on model-building than the requirement of producing $m_s$ at that scale.
 Subsequently, we quantify the amount
of technicolor condensate enhancement required to produce a given quark mass if
the ETC scale is of order $10^3$ TeV. Our quantitative results are intended to provide
a framework within which to interpret and apply lattice Monte Carlo studies of candidate
walking technicolor theories, 
such as those in Refs. \cite{Catterall:2007yx,Catterall:2008qk,Appelquist:2009ka,Deuzeman:2009mh,Fodor:2009wk,Nagai:2009ip,Svetitsky:2009pz,Bursa:2009we,DelDebbio:2010hx,DeGrand:2009hu,Hasenfratz:2010fi,DeGrand:2010na}.


\section{Constraints on $\Lambda_{ETC}$ from neutral meson mixing}\label{sec:FCNC}

At low energies, the flavor-changing four-fermion interactions induced by ETC boson exchange alter the predicted rate of neutral meson mixing.  Ref. \cite{Bona:2007vi} has derived constraints on general $\Delta F = 2$ four-fermion operators that affect neutral Kaon, D-meson, and B-meson mixing, including the effects of running from the new physics scale down to the meson scale and interpolating between quark and meson degrees of freedom.   Their limits on the coefficients ($C^1_j$) of the FCNC operators involving LH current-current interactions:
\begin{eqnarray}
C^1_K (\bar{s}_L \gamma^\mu d_L) (\bar{s}_L \gamma_\mu d_L) \label{eq:cdeffs}\\
C^1_D\, (\bar{c}_L \gamma^\mu u_L) (\bar{c}_L \gamma_\mu u_L) \\
C^1_{B_d} (\bar{b}_L \gamma^\mu d_L) (\bar{b}_L \gamma_\mu d_L) \\
\! C^1_{B_s} (\bar{b}_L \gamma^\mu s_L) (\bar{b}_L \gamma_\mu s_L) \,,\!
\end{eqnarray}
are listed in the left column of Table \ref{tab:one}.  In their notation, the generic form of the coefficient $C^1_i$ is:
\begin{equation}
C^1_i = \frac{F_i L_i}{\Lambda^2}
\end{equation}
where $F_i$ is a flavor factor that is expected to be $\vert F_i\vert \sim 1$ in a model with an arbitrary flavor structure
and absent any GIM-like mechanism 
\cite{Chivukula:1987py,D'Ambrosio:2002ex,Appelquist:2004ai,Martin:2004ec} for suppressing 
flavor-changing neutral currents; $L_i$ is a loop factor that is simply $1$ in a model with tree-level FCNC; and $\Lambda$ is the scale of new physics. 

\begin{table}[h]
\caption{Limits from the UTFit Collaboration \cite{Bona:2007vi} on coefficients of left-handed four-fermion operators contributing to neutral meson mixing (left column) and the implied lower bound on the ETC scale (right column).  The bounds in the first four rows apply when one assumes ETC does not contribute to CP violation; the bound in the last row applies if one assumes that ETC does contribute to CP violation in the Kaon system. }
\begin{center}
\begin{tabular}{|c|c|} \hline
$\phantom{\dfrac{\strut}{\strut}} $\ \ Bound on operator coefficient\  (GeV$^{-2}$)\ \ &  \ \ Implied lower limit on ETC scale \ ($10^3$ TeV)\ \  \\  \hline\hline
$\phantom{\dfrac{\strut}{\strut}} - 9.6 \times 10^{-13} < \Re(C^1_K) < 9.6 \times 10^{-13}$  & $1.0$ \\  \hline 
$\phantom{\dfrac{\strut}{\strut}}  \vert C^1_D \vert  < 7.2 \times 10^{-13}$ & $1.5$ \\  \hline
$\phantom{\dfrac{\strut}{\strut}}\vert C^1_{B_d} \vert  < 2.3 \times 10^{-11}$ & $0.21$ \\ \hline
$\phantom{\dfrac{\strut}{\strut}}\vert C^1_{B_s} \vert  < 1.1 \times 10^{-9}$ & $0.03$\\  \hline \hline
$\phantom{\dfrac{\strut}{\strut}}- 4.4 \times 10^{-15} < \Im(C^1_K) < 2.8 \times 10^{-15}$ & $10$ \\  \hline
\end{tabular}
\end{center}
\label{tab:one}
\end{table}

 In the case of an ETC model with arbitrary flavor structure and no assumed ETC contribution to CP-violation,  one has  $C^1_i = \Lambda_{ETC}^{-2}$ and the limits on the $\Lambda_{ETC}$ from \cite{Bona:2007vi} are as shown in the right-hand column of Table \ref{tab:one}.  The lower bound on $\Lambda_{ETC}$ from $D$-meson mixing is now the strongest, with that from Kaon mixing a close second and those from $B$-meson mixing far weaker.  Since the charm quark is so much heavier than the strange quark, requiring an ETC model to produce $m_c$ from interactions at a scale of over $1000$ TeV is a significantly stronger constraint on model-building than the requirement of producing $m_s$ at that scale.
Note that if one, instead, assumes that ETC contributes to CP-violation in the Kaon system, then the relevant bound on $\Lambda_{ETC}$ comes from the imaginary part of $C^1_K$  and is a factor of ten more severe (see last row of Table \ref{tab:one}).

In the next section we will quantify the amount
of technicolor condensate enhancement required to produce a given quark mass if
the ETC scale is of order $10^3$ TeV.


\section{Condensate Enhancement and $\gamma_m$}\label{sec:gammam}

In studying how ETC theories produce quark masses, the primary operator of interest has the form\footnote{In an ETC gauge theory, we would expect
$1/\Lambda^2_{ETC} \equiv g^2_{ETC}/M^2_{ETC}$ where $g_{ETC}$ and $M_{ETC}$ are
the appropriate extended technicolor coupling and gauge-boson mass, respectively. At energies below
$M_{ETC}$, these parameters always appear (to leading order in the ETC interactions) in this
ratio -- and therefore, we use $\Lambda_{ETC}$ for simplicity.}
\begin{equation}
\frac{(\bar{Q}^a_L \gamma^\mu q^j_L)(u^i_R \gamma_\mu U^a_R)}{\Lambda^2_{ETC}}~,
\end{equation}
where the $Q^a_L$ and $U^a_R$ are technifermions ($a$ is a technicolor index), and the
$q^j_L$ and $u^i_R$ are left-handed quark doublet and right-handed up-quark gauge-eigenstate
fields ($i$ and $j$ are family indices). This operator will give rise, after technifermion chiral symmetry breaking at
the weak scale, to a fermion mass term of order 
\begin{equation}
{\cal M}_{ij} = \frac{\langle \bar{U}_L U_R \rangle_{\Lambda_{ETC}}}{\Lambda^2_{ETC}}~.
\label{eq:i}
\end{equation}
Here it is important to note that the technifermion condensate, $\langle \bar{U}_L U_R \rangle_{\Lambda_{ETC}}$
is renormalized at the ETC scale \cite{Holdom:1981rm,Holdom:1984sk,Yamawaki:1985zg,Appelquist:1986an,Appelquist:1986tr,Appelquist:1987fc}. It is related to the condensate at the technicolor (electroweak symmetry
breaking) scale by
\begin{equation}
\langle \bar{U}_L U_R \rangle_{\Lambda_{ETC}} = 
\exp\left(\int_{\Lambda_{TC}}^{\Lambda_{ETC}}\gamma_m(\alpha_{TC}(\mu))\frac{d\mu}{\mu}\right)
\langle \bar{U}_L U_R \rangle_{\Lambda_{TC}}~,
\label{eq:ii}
\end{equation}
where $\gamma_m(\alpha_{TC}(\mu))$ is the anomalous dimension of the technifermion mass operator.\footnote{We will
address the potential scheme-dependence of $\gamma_m$ below.}
Using an estimate of the technifermion condensate, and a calculation of the anomalous dimension of
the mass operator, we may estimate the size of quark mass which can arise in a technicolor theory for a
given ETC scale.

In a theory of walking technicolor \cite{Holdom:1981rm,Holdom:1984sk,Yamawaki:1985zg,Appelquist:1986an,Appelquist:1986tr,Appelquist:1987fc}, the gauge coupling runs very slowly just above the technicolor scale
$\Lambda_{TC}$. The largest enhancement occurs in the limit of ``extreme walking"
in which the technicolor coupling, and hence the anomalous dimension $\gamma_m$, remains
approximately constant from the technicolor scale, $\Lambda_{TC}$, all the way to the ETC scale,
$\Lambda_{ETC}$. In the limit of extreme walking, one obtains
\begin{equation}
\langle \bar{U}_L U_R \rangle_{\Lambda_{ETC}} =
\left(\frac{\Lambda_{ETC}}{\Lambda_{TC}}\right)^{\gamma_m}
\langle \bar{U}_L U_R \rangle_{\Lambda_{TC}}~.
\label{eqn:enhance}
\end{equation}

We may now use (\ref{eqn:enhance}) to quantify the enhancement of the technicolor condensate required to produce the observed quark masses in a walking model.  Specifically, we will investigate the size of the quark mass which can be achieved in the limit of extreme walking for various $\gamma_m$, and an ETC scale of $10^3$ TeV (which, as shown above, should suffice to meet the CP-conserving
FCNC constraints in the $K$ - and $D$-meson systems).   The calculation requires an estimate of the technicolor scale $\Lambda_{TC}$ and  the technicolor condensate renormalized at the electroweak scale, $\langle \bar{U}_L U_R \rangle_{\Lambda_{TC}}$.

Two estimates of the scales associated with technicolor chiral symmetry breaking are commonly
used in the literature: Naive Dimensional Analysis (NDA) \cite{Weinberg:1978kz,Manohar:1983md,Georgi:1985hf}
and simple dimensional analysis (DA) as applied in \cite{Eichten:1979ah}. In Naive Dimensional Analysis, one
associates $\Lambda_{TC}$ with the ``chiral symmetry breaking scale" for the technicolor
theory, and hence $\Lambda_{TC}=\Lambda_{\chi SB}\approx4\pi v$ (where $v\approx 250$ GeV is the analog of $f_\pi$
in QCD), while
\begin{equation}
\langle \bar{U}_L U_R \rangle_{\Lambda_{TC}} \approx \frac{\Lambda^3_{\chi SB}}{(4\pi)^2}~.
\end{equation}
Inserting these relations into eqns. (\ref{eq:i}) and (\ref{eq:ii}) we find, in the limit
of extreme walking (constant $\gamma_m$)
\begin{eqnarray}
m^{NDA}_q &=& \frac{\Lambda_{\chi SB}}{(4\pi)^2}
\left(\frac{\Lambda_{\chi SB}}{\Lambda_{ETC}}\right)^{2-\gamma_m}\\
&\approx& 19.9\, {\rm GeV} \cdot \,(3.14 \times 10^{3})^{2-\gamma_m}~, \label{eq:iii}
\end{eqnarray}
where the last equality applies for an ETC scale of $10^3$ TeV.  Alternatively, in
the simple dimensional estimates given for example in \cite{Eichten:1979ah},
one simply assumes that all technicolor scales are given by $\Lambda_{TC} \approx
1$ TeV, and hence one uses
\begin{equation}
\langle \bar{U}_L U_R \rangle_{\Lambda_{TC}} \approx \Lambda^3_{TC}.
\end{equation}
In this case, one finds 
\begin{eqnarray}
m^{DA}_q & = & \Lambda_{TC} \left(\frac{\Lambda_{TC}}{\Lambda_{ETC}}\right)^{2-\gamma_m}\\
& \approx & 1000\,{\rm GeV} \cdot \left(1.0\times 10^{-3}\right)^{2-\gamma_m}~, \label{eq:iv}
\end{eqnarray}
where, again, the last equality applies for an ETC scale of $10^3$ TeV. 
Note that in neither case have we 
included factors to correct for the number of weak doublets in the technicolor sector, nor
attempted to account for the ``large-$N_{TC}$" limit \cite{Hill:2002ap} -- however, such factors
can only {\it suppress} the size of the quark masses generated.

In Table \ref{tab:two} we use eqns. (\ref{eq:iii}) and (\ref{eq:iv}) to estimate the size of
quark mass corresponding to various (constant) values of $\gamma_m$ and an ETC
scale of $10^3$ TeV. We show these values in the range $0 \le \gamma_m \le 2.0$
since $\gamma_m\simeq 0$ in a ``running" technicolor theory, and  conformal group representation unitarity implies 
that $\gamma_m \le 2.0$ \cite{Mack:1975je}. 
The usual Schwinger-Dyson analysis used to analyze technicolor theories would
imply that $\gamma_m \le 1.0$ in walking technicolor theories \cite{Holdom:1981rm,Holdom:1984sk,Yamawaki:1985zg,Appelquist:1986an,Appelquist:1986tr,Appelquist:1987fc}, while the values $1.0 \le
\gamma_m \le 2.0$
could occur in strong-ETC theories \cite{Appelquist:1988as,Takeuchi:1989qa,Miransky:1988gk,Fukano:2010yv,Matumoto:1989hf}.

\begin{table}[h]
\caption{Size of the quark mass $m_q$ generated by technicolor dynamics assuming an ETC scale $\Lambda_{ETC} = 1000$ TeV and various values for the anomalous dimension $\gamma_m$ of the mass operator. In the row labeled NDA [DA], the value of the techniquark condensate at the technicolor scale is taken to be $\langle \bar{T} T \rangle \approx (580\, {\rm GeV})^3$ [$(1000\, {\rm GeV})^3$].  Values of $\gamma_m$ of 1.0 or less correspond to walking theories
\protect\cite{Holdom:1981rm,Holdom:1984sk,Yamawaki:1985zg,Appelquist:1986an,Appelquist:1986tr,Appelquist:1987fc}; values greater than 1.0 correspond to strong-ETC theories \protect\cite{Appelquist:1988as,Takeuchi:1989qa,Miransky:1988gk,Fukano:2010yv,Matumoto:1989hf}.}
\begin{center}
\begin{tabular}{|c||c|c|c|c|c||c|c|c|c|} \hline
$\phantom{\dfrac{\strut}{\strut}} \gamma_m$ &0 & 0.25 & 0.5 & 0.75 & 1.0 & 1.25& 1.5 & 1.75 & 2.0 \\  \hline \hline
$\phantom{\dfrac{\strut}{\strut}} m^{NDA}_q$  \ \ & \ \  0.2 MeV \ \ & \ \ 0.8 MeV \ \ & \ \ 3.5 MeV \ \ & \ \ 15 MeV \ \ & \ \ 63 MeV \ \ & \ \ 260 MeV \ \ & \ \ 1.1 GeV \ \ & \ \ 4.7 GeV \ \ & \ \ 20 GeV\\  \hline 
$\phantom{\dfrac{\strut}{\strut}} m^{DA}_q$  \ \ & \ \  1 MeV \ \ & \ \ 5.6 MeV \ \ & \ \ 32 MeV \ \ & \ \ 180 MeV \ \ & \ \ 1 GeV \ \ & \ \ 5.6 GeV \ \ & \ \ 32 GeV \ \ & \ \ 180 GeV \ \ & \ \ 1 TeV\\  \hline 

\end{tabular}
\end{center}
\label{tab:two}
\end{table}


\section{Discussion}\label{sec:discussion}

Examining Table \ref{tab:two}, we see that generating the charm quark mass from
ETC dynamics at a scale of order $10^3$ TeV requires an anomalous dimension $\gamma_m$
close to or exceeding one, even in the case of the more generous DA estimate of the
technifermion condensate.  It will therefore be helpful for nonpertubative studies of strong technicolor dynamics to determine how large $\gamma_m$ can be in specific candidate theories of walking technicolor. Values of $\gamma_m$ substantially less than one would require a lower ETC scale, which would necessitate the construction of ETC theories with approximate flavor symmetries \cite{Chivukula:1987py,D'Ambrosio:2002ex,Appelquist:2004ai,Martin:2004ec} and 
corresponding GIM-like partial cancellations of flavor-changing contributions. Note also that our quark mass estimates are generous on several fronts: taking into account the number of weak doublets in the technicolor sector, large $N_{TC}$ effects, or less extreme walking of the technicolor coupling would suppress the size of the quark mass generated. 

Our results further suggest 
that the nonperturbative study of strong-ETC models \cite{Appelquist:1988as,Takeuchi:1989qa,Miransky:1988gk,Fukano:2010yv,Matumoto:1989hf}
may also be useful, since generating the heavy quark masses may be easier in such models. In this case, it will be particularly interesting to see if such theories contain
a light, but broad, scalar Higgs-like resonance and whether they could avoid potentially
dangerous custodial symmetry violating contributions to $M^2_W/M^2_Z$
 \cite{Chivukula:1990bc,Appelquist:1991kn,Carone:1992rh}.

Finally we should address a subtlety in our discussion: $\gamma_m$ is only scheme-independent at an IR 
fixed point where the gauge theory is conformal. In fact, lattice Monte Carlo studies to date
\cite{Catterall:2007yx,Catterall:2008qk,Appelquist:2009ka,Deuzeman:2009mh,Fodor:2009wk,Nagai:2009ip,Svetitsky:2009pz,Bursa:2009we,DelDebbio:2010hx,DeGrand:2009hu,Hasenfratz:2010fi,DeGrand:2010na}  focus
on establishing the ``conformal window" of strongly coupled theories within which, since the
theory is truly conformal, chiral symmetry breaking (and therefore electroweak symmetry
breaking) would not occur. 
As we discuss above, however, candidate walking-technicolor theories will likely be very close to conformal over a 
large range of energy scales -- namely, they are expected to be approximately conformal over the three orders
of magnitude that separate the technicolor and ETC scales. Therefore, measurements of
$\gamma_m$ in the conformal phase of these theories can suggest (via the results of Table \ref{tab:two}) 
which models might form the basis of a realistic technicolor
model, when either ``tuned" to be slightly away from the fixed point trajectory,
or deformed \cite{Luty:2004ye} by the presence of some additional operator (see also the
discussion of the utility of working in the conformal phase in Ref. \cite{DeGrand:2009hu}). Ultimately, it would be desirable to simulate a nearly conformal walking-technicolor theory
in the phase of broken chiral symmetry -- in which case the relevant technicolor condensate and 
ETC-generated ordinary fermion masses can be measured directly.

In this paper we have noted that constraints on FCNC in the neutral $D$-meson 
system imply that the charm quark mass must, like the strange quark mass, arise from flavor dynamics at a scale 
of order $10^3$ TeV. We have also quantified the degree to which the technicolor condensate must be 
enhanced in order to yield the observed quark masses, if the extended technicolor scale 
is of order $10^3$ TeV. Our results provide a framework in which to interpret
and apply the results of lattice studies of conformal strongly interacting gauge theories, 
and the corresponding numerical 
measurements of the anomalous dimension of the mass operator in candidate 
theories of walking technicolor.


\section{Acknowledgements}

This work was supported in part by the US National Science Foundation under grants PHY-0354226
and PHY-0854889.  RSC and EHS gratefully acknowledge conversations with Tom Appelquist, Tom DeGrand, 
Luigi Del Debbio, Francesco Sannino, 
Bob Shrock, Ben Svetitsky, and Rohana Wijewhardana, as well as participants in the
``Strong Coupling Beyond the Standard Model" workshop held at the Aspen Center for Physics
during May and June, 2010. We also thank the Aspen Center for Physics for its support while
this work was completed.


\begin{thebibliography}{199}


\bibitem{Weinberg:1979bn}
  S.~Weinberg,
  Phys.\ Rev.\  D {\bf 19}, 1277 (1979).

\bibitem{Susskind:1978ms}
  L.~Susskind,
  Phys.\ Rev.\  D {\bf 20}, 2619 (1979).

\bibitem{Hill:2002ap}
For a reviews, see  C.~T.~Hill and E.~H.~Simmons,
  Phys.\ Rept.\  {\bf 381}, 235 (2003)
  [Erratum-ibid.\  {\bf 390}, 553 (2004)]
  [arXiv:hep-ph/0203079] and
  R.~S.~Chivukula, M.~Narain and J.~Womersley,
  pages 1258-1264
  of  C.~Amsler {\it et al.}  [Particle Data Group],
  Phys.\ Lett.\  B {\bf 667}, 1 (2008).
  
\bibitem{Dimopoulos:1979es}
  S.~Dimopoulos and L.~Susskind,
  Nucl.\ Phys.\  B {\bf 155}, 237 (1979).

\bibitem{Eichten:1979ah}
  E.~Eichten and K.~D.~Lane,
  Phys.\ Lett.\  B {\bf 90}, 125 (1980).
  
\bibitem{Chivukula:1987py}
  R.~S.~Chivukula and H.~Georgi,
  Phys.\ Lett.\  B {\bf 188}, 99 (1987).

\bibitem{D'Ambrosio:2002ex}
  G.~D'Ambrosio, G.~F.~Giudice, G.~Isidori and A.~Strumia,
  Nucl.\ Phys.\  B {\bf 645}, 155 (2002)
  [arXiv:hep-ph/0207036].

\bibitem{Appelquist:2004ai}
For a recent discussion of approximate flavor symmetries
in ETC models, see  T.~Appelquist, N.~D.~Christensen, M.~Piai and R.~Shrock,
  Phys.\ Rev.\  D {\bf 70}, 093010 (2004)
  [arXiv:hep-ph/0409035]. In the class of theories discussed in this
  paper (and references therein), simultaneously
  avoiding {\it both} $D$- and $K$-meson mixing constraints can be difficult.

\bibitem{Martin:2004ec}
  A.~Martin and K.~Lane,
  Phys.\ Rev.\  D {\bf 71}, 015011 (2005)
  [arXiv:hep-ph/0404107].

\bibitem{Holdom:1981rm}
  B.~Holdom,
  Phys.\ Rev.\  D {\bf 24}, 1441 (1981).

\bibitem{Holdom:1984sk}
  B.~Holdom,
  Phys.\ Lett.\  B {\bf 150}, 301 (1985).

\bibitem{Yamawaki:1985zg}
  K.~Yamawaki, M.~Bando and K.~i.~Matumoto,
  Phys.\ Rev.\ Lett.\  {\bf 56}, 1335 (1986).

\bibitem{Appelquist:1986an}
  T.~W.~Appelquist, D.~Karabali and L.~C.~R.~Wijewardhana,
  Phys.\ Rev.\ Lett.\  {\bf 57}, 957 (1986).


\bibitem{Appelquist:1986tr}
  T.~Appelquist and L.~C.~R.~Wijewardhana,
  Phys.\ Rev.\  D {\bf 35}, 774 (1987).

\bibitem{Appelquist:1987fc}
  T.~Appelquist and L.~C.~R.~Wijewardhana,
  Phys.\ Rev.\  D {\bf 36}, 568 (1987).

\bibitem{Appelquist:1997fp}
  T.~Appelquist, J.~Terning and L.~C.~R.~Wijewardhana,
  Phys.\ Rev.\ Lett.\  {\bf 79}, 2767 (1997)
  [arXiv:hep-ph/9706238].

\bibitem{Sannino:2009za}
  F.~Sannino,
  arXiv:0911.0931 [hep-ph].


\bibitem{Miransky:1988gk}
  V.~A.~Miransky and K.~Yamawaki,
  Mod.\ Phys.\ Lett.\  A {\bf 4}, 129 (1989).

\bibitem{Matumoto:1989hf}
  K.~Matumoto,
  Prog.\ Theor.\ Phys.\  {\bf 81}, 277 (1989).

\bibitem{Appelquist:1988as}
  T.~Appelquist, M.~Einhorn, T.~Takeuchi and L.~C.~R.~Wijewardhana,
  Phys.\ Lett.\  B {\bf 220}, 223 (1989).

\bibitem{Takeuchi:1989qa}
  T.~Takeuchi,
  Phys.\ Rev.\  D {\bf 40}, 2697 (1989).

\bibitem{Fukano:2010yv}
For a recent analysis, see  H.~S.~Fukano and F.~Sannino,
  arXiv:1005.3340 [hep-ph].


\bibitem{Peskin:1990zt}
  M.~E.~Peskin and T.~Takeuchi,
  Phys.\ Rev.\ Lett.\  {\bf 65}, 964 (1990).

\bibitem{Peskin:1991sw}
  M.~E.~Peskin and T.~Takeuchi,
  Phys.\ Rev.\  D {\bf 46}, 381 (1992).


\bibitem{Catterall:2007yx}
  S.~Catterall and F.~Sannino,
  Phys.\ Rev.\  D {\bf 76}, 034504 (2007)
  [arXiv:0705.1664 [hep-lat]].

\bibitem{Catterall:2008qk}
  S.~Catterall, J.~Giedt, F.~Sannino and J.~Schneible,
  JHEP {\bf 0811}, 009 (2008)
  [arXiv:0807.0792 [hep-lat]].

\bibitem{Svetitsky:2009pz}
  B.~Svetitsky,
  Nucl.\ Phys.\  A {\bf 827}, 547C (2009)
  [arXiv:0901.2103 [hep-lat]].


\bibitem{Deuzeman:2009mh}
  A.~Deuzeman, M.~P.~Lombardo and E.~Pallante,
  arXiv:0904.4662 [hep-ph].

\bibitem{Fodor:2009wk}
  Z.~Fodor, K.~Holland, J.~Kuti, D.~Nogradi and C.~Schroeder,
  Phys.\ Lett.\  B {\bf 681}, 353 (2009)
  [arXiv:0907.4562 [hep-lat]].

\bibitem{Nagai:2009ip}
  K.~i.~Nagai, G.~Carrillo-Ruiz, G.~Koleva and R.~Lewis,
  Phys.\ Rev.\  D {\bf 80}, 074508 (2009)
  [arXiv:0908.0166 [hep-lat]].


\bibitem{Appelquist:2009ka}
  T.~Appelquist {\it et al.},
  Phys.\ Rev.\ Lett.\  {\bf 104}, 071601 (2010)
  [arXiv:0910.2224 [hep-ph]].

\bibitem{DeGrand:2009hu}
 T.~DeGrand,
 Phys.\ Rev.\  D {\bf 80}, 114507 (2009)
 [arXiv:0910.3072 [hep-lat]].


\bibitem{Bursa:2009we}
 F.~Bursa, L.~Del Debbio, L.~Keegan, C.~Pica and T.~Pickup,
 Phys.\ Rev.\  D {\bf 81}, 014505 (2010)
 [arXiv:0910.4535 [hep-ph]].

\bibitem{DelDebbio:2010hx}
  L.~Del Debbio, B.~Lucini, A.~Patella, C.~Pica and A.~Rago,
  arXiv:1004.3206 [hep-lat], and references therein.

\bibitem{Hasenfratz:2010fi}
 A.~Hasenfratz,
 arXiv:1004.1004 [hep-lat].

\bibitem{DeGrand:2010na}
  T.~DeGrand, Y.~Shamir and B.~Svetitsky,
  arXiv:1006.0707 [hep-lat].

\bibitem{Bona:2007vi}
  M.~Bona {\it et al.}  [UTfit Collaboration],
  JHEP {\bf 0803}, 049 (2008)
  [arXiv:0707.0636 [hep-ph]].


\bibitem{Weinberg:1978kz}
  S.~Weinberg,
  Physica A {\bf 96}, 327 (1979).

\bibitem{Manohar:1983md}
  A.~Manohar and H.~Georgi,
  Nucl.\ Phys.\  B {\bf 234}, 189 (1984).

\bibitem{Georgi:1985hf}
  H.~Georgi,
  Nucl.\ Phys.\  B {\bf 266}, 274 (1986).

\bibitem{Mack:1975je}
  G.~Mack,
  Commun.\ Math.\ Phys.\  {\bf 55}, 1 (1977).


\bibitem{Chivukula:1990bc}
  R.~S.~Chivukula, A.~G.~Cohen and K.~D.~Lane,
  Nucl.\ Phys.\  B {\bf 343}, 554 (1990).

\bibitem{Appelquist:1991kn}
  T.~Appelquist, J.~Terning and L.~C.~R.~Wijewardhana,
  Phys.\ Rev.\  D {\bf 44}, 871 (1991).

\bibitem{Carone:1992rh}
  C.~D.~Carone and E.~H.~Simmons,
  Nucl.\ Phys.\  B {\bf 397}, 591 (1993)
  [arXiv:hep-ph/9207273].

\bibitem{Luty:2004ye}
  M.~A.~Luty and T.~Okui,
  JHEP {\bf 0609}, 070 (2006)
  [arXiv:hep-ph/0409274].


\end{thebibliography}
\end{document}